\documentclass[10pt,tightenlines,eqsecnum,floats,aps,nofootinbib,prd]{revtex4-2}
\usepackage{amsmath,amssymb,amsfonts,amsthm,amscd}
\usepackage{graphicx}
\usepackage{amsmath}
\usepackage{amssymb}
\usepackage{graphicx}
\usepackage{cancel}
\usepackage{slashed}
\usepackage{mathtools}
\usepackage{tcolorbox}
\usepackage{tikz-cd}
\usepackage{tikz}
\usepackage{mathrsfs}
\usepackage{relsize}
\usepackage{natbib}
\usepackage[margin=1in]{geometry}
\usepackage{hyperref}
\usepackage[normalem]{ulem}

\usetikzlibrary{shapes.geometric, arrows}
\tikzstyle{startstop} = [rectangle, rounded corners, minimum width=3cm, minimum height=1cm,text centered, draw=black, fill=red!30]
\tikzstyle{arrow} = [thick,->,>=stealth]
\begin{document}
\title{Dynamical Similarity in Multisymplectic Field Theory}
\author{Callum Bell}
\email{c.bell8@lancaster.ac.uk}
\affiliation{Department of Physics, Lancaster University, Lancaster UK}

\author{David Sloan}
\email{d.sloan@lancaster.ac.uk}
\affiliation{Department of Physics, Lancaster University, Lancaster UK}

\begin{abstract}
    Symmetry under a particular class of non-strictly canonical transformation may be used to identify, and subsequently excise degrees of freedom which do not contribute to the closure of the algebra of dynamical observables. Such redundant degrees of freedom may physically be identified with empirically-inaccessible measures of global scale. In this article, we present a mathematical framework which extends the symmetry reduction procedure to theories of classical fields, in both the Lagrangian and Hamiltonian settings. In order to maintain Lorentz covariance, while simultaneously working with a finite-dimensional phase space, we employ the De Donder-Weyl formalism, for which the natural description is formulated in terms of the fibered manifolds of multisymplectic geometry. We subsequently analyse a number of simple examples, and provide a discussion of the broader implications of our construction. 
\end{abstract}
\maketitle
\section{Introduction}\label{Sec:Introduction}
Our most precise and well-tested description of nature is currently formulated in the language of field theory, both at a classical and quantum level \cite{nair2006quantum}. Indeed, the advent of a field-theoretic viewpoint has arguably been one of the most significant shifts in perspective of the last few centuries. The particular mathematical framework introduced to model the phenomena of interest depends upon context, and, to a lesser extent, convenience and personal preference. From the perspective of the symplectic description of classical field theories, we highlight two philosophically-distinct approaches. The canonical formalism, often employed to construct the quantum theory from the underlying classical framework, studies dynamical evolution, as defined by the initial-value problem. In particular, one sacrifices Lorentz covariance, selecting a privileged time coordinate, and considers field variables defined on spatial slices of fixed temporal parameter \cite{sardanashvily1995generalized}.\\

The spacetime-covariant formalism, by contrast, treats events at a single spacetime point, considering all coordinates (including time) equally \cite{aldaya1980geometric,crnkovic1987covariant}. Multisymplectic geometry provides a rigorous framework within which to study classical field theories in just such a covariant setting, which we shall refer to as the De Donder-Weyl formalism \cite{de1930theorie,weyl1935geodesic,klusovn2023weyl,vasak2024ccgg,gieres2023covariant}. While the canonical approach, or alternative tools, such as variational bicomplexes, are often favoured in the literature \cite{anderson1992introduction}, their infinite-dimensionality presents a significant hindrance to the study of the class of symmetries of interest in the present work \cite{kanatchikov1994basic}, which we shall refer to as \textit{scaling symmetries} or \textit{dynamical similarities}.\\

Scaling symmetries arise when our mathematical formalism contains more structure than that required to describe the dynamical evolution of the observable degrees of freedom \cite{sloan2018dynamical}. Not only is such a scenario undesirable from the perspective of elegance and simplicity, but the superfluous structure can, at times, introduce pathological behaviour that is an artifact of our choice of description, and \textit{not} reflective of the true nature of the underlying physical system. For example, it is the universally-accepted practice to describe the dynamical evolution of homogeneous, isotropic cosmologies using a dimensionless scale factor $a(t)$ \cite{wald2010general}. It is known that the value of the scale factor by itself cannot be determined by experiment; it is only the \textit{ratio} of $a(t)$ at two different times that encodes physical information. It has been demonstrated that one can formulate an equivalent cosmological description, without ever introducing a scale factor \cite{sloan2021new,sloan2023herglotz}. Moreover, as alluded to above, this description is perfectly well-defined at those points where the singular nature of the scale factor causes the conventional theory to break down \cite{hoffmann2024continuation}.\\

We argue that, following the Principle of Essential and Sufficient Autonomy, when seeking a theoretical description of the natural world, one should first construct a framework which amply describes the phenomena of interest, before progressively reducing its superfluous structure, until a minimally-sufficient yet consistent theory remains \cite{gryb2021scale,ismael2021symmetry,ismael2023rethinking,vanFraassen_2012}. By considering the multisymplectic formulation of classical field theory, introduced in section (\ref{Sec:MultisymplecticFieldTheory}), we demonstrate that invariance under changes to a system's global scale indicates the presence of a single redundant degree of freedom, corresponding to the generator of these rescalings. The process by which such structure is excised from our theory is known as `contact reduction', and is introduced in section (\ref{Sec:DynamicalSimilarity}). We then dedicate sections (\ref{Sec:ContactReductionLagrangian}) and (\ref{Sec:ContactReductionHamiltonian}) to the field-theoretic generalisation of this procedure. Throughout, the multisymplectic framework is indispensable for our construction in two regards; on the one hand, it is not clear that the more conventional approaches admit a means to assign a well-defined structure to our theory, after having eliminated a scaling degree of freedom. By contrast, when the fibered manifolds of the multisymplectic formalism are employed, we shall demonstrate that the reduced theory is defined on a space naturally identified as a multicontact manifold \cite{de2023multicontact,de2025practical,rivas2025symmetries}. The second way in which the multisymplectic framework proves highly fruitful is its finite-dimensionality. In working with a manifestly finite-dimensional (velocity) phase space, we are able to very easily perform a counting of the degrees of freedom, showing that the reduction procedure eliminates precisely one.\\

The ideas presented throughout constitute a necessary preliminary framework that is precursory to a full treatment of contact-reduced gauge theories. Our principal objective in developing methods for treating singular theories lies in the study of General Relativity, where it is known that the Einstein-Hilbert action admits a scaling symmetry \cite{sloan2025dynamical}. A geometrical constraint algorithm has been developed for singular field theories described using multisymplectic geometry \cite{de1996geometrical,de2005pre}, and the interaction between phase space restriction, and the contact reduction procedure presented here should be examined, prior to analysing scaling symmetries of gravitational actions.\\

Finally, we close with a number of examples, in which a scaling symmetry is identified, and the reduction procedure carried out. These examples have been chosen to be sufficiently involved so as to admit a detailed discussion of the physical implications of our construction. They are, however, simple enough so as to allow readers unfamiliar with multisymplectic geometry to follow, without additional contextual complications. Having presented these examples, we shall indicate a number of more physically-motivated theories to which our formalism is applicable, as well as discussing several open questions and lines of further investigation.
\section{Multisymplectic Formulation of Classical Field Theories}\label{Sec:MultisymplecticFieldTheory}
Our treatment of multisymplectic geometry aims to be self-contained, but by no means exhaustive; since our objective is to describe classical field theories in a covariant manner, much of the material presented throughout this section is adapted with this goal in mind. A more complete and in-depth discussion of the ideas discussed may be found in \cite{cantrijn1996hamiltonian,cantrijn1999geometry,forger2013multisymplectic}, for example.
\subsection{Multisymplectic Manifolds and Multivector Fields}\label{Subsec:MultisymplecticManifolds}
In general, we say that an $m$-dimensional differentiable manifold $\mathcal{M}$ is multisymplectic if it admits a closed, 1-non-degenerate $k$-form $\Omega\in\Omega^k(\mathcal{M})$, with $1<k\leqslant m$ \cite{roman2019some}. That $\Omega$ is 1-non-degenerate means that for every $p\in \mathcal{M}$ and $X_p\in T_p\mathcal{M}$
\begin{equation*}
    \iota_{X_p}\Omega_p = 0 \quad\quad\implies \quad\quad X_p=0
\end{equation*}
If $\Omega$ is closed but 1-degenerate, we refer to the pair $(\mathcal{M},\Omega)$ as a pre-multisymplectic manifold. Given a multisymplectic manifold $(\mathcal{M},\Omega)$, sections of the $k^{\textrm{th}}$ exterior power of $T\mathcal{M}$ define a class of objects known as \textit{multivector fields of degree $k$} on $\mathcal{M}$. We denote the space of all such multivector fields $\mathfrak{X}^k(\mathcal{M}):= \Gamma(\mathcal{M},\wedge^k\,T\mathcal{M})$. A $k$-multivector field $\boldsymbol{X}\in\mathfrak{X}^k(\mathcal{M})$ is said to be \textit{locally decomposable} if, for every point $p\in\mathcal{M}$, there exists an open neighbourhood $\mathcal{U}_p\subset \mathcal{M}$, and vector fields $X_1,\,\cdots\,,X_k\in \mathfrak{X}^{\infty}(\mathcal{U}_p)$, such that
\begin{equation*}
    \boldsymbol{X}|_{\mathcal{U}_p} = X_1 \wedge\,\cdots\,\wedge X_k
\end{equation*}
We say that an $m$-dimensional distribution $\mathcal{D}\subset T\mathcal{M}$ is \textit{locally associated} to a non-zero $\boldsymbol{X}\in \mathfrak{X}^m(\mathcal{M})$, if there exists some connected open set $\mathcal{V}\subset \mathcal{M}$, such that $\boldsymbol{X}|_{\mathcal{V}}\in \Gamma(\mathcal{V},\wedge^m\mathcal{D}|_{\mathcal{V}})$, and $\boldsymbol{X}\in \mathfrak{X}^m(\mathcal{M})$ is said to be \textit{integrable} if its locally associated distribution is integrable.\\

Finally, we define the contraction between a locally decomposable $m$-multivector field $\boldsymbol{X}=X_1\wedge\,\cdots\,\wedge X_m$ and a differential $k$-form $\Xi \in \Omega^k(\mathcal{M})$ as
\begin{equation*}
    \iota_{\scriptscriptstyle\boldsymbol{X}} \Xi = \begin{cases}
        \iota_{\scriptscriptstyle X_m}\,\cdots\;\iota_{\scriptscriptstyle X_1}\,\Xi \quad &\textrm{if $m\leqslant k$}\\
        0 &\textrm{if $m>k$}
    \end{cases}
\end{equation*}
\subsection{Lagrangian Field Theory}\label{Subsec:LagrangianFieldTheory}
We begin by introducing a fibre bundle $\pi:E\rightarrow M$ over the orientable, $d$-dimensional spacetime manifold $M$, with volume form $\omega$, and local coordinates $x^{\mu}$ ($\mu=0,\,\cdots,d-1$). The total space $E$ is a $(d+n)$-dimensional manifold, referred to as the \textit{covariant configuration space}, upon which local coordinates are written $(x^{\mu},y^a)$, with $1\leqslant a \leqslant n$. Introducing the first jet bundle $\kappa:J^1E\rightarrow E$ of sections of $\pi$ \cite{saunders1989geometry,kupershmidt2006geometry}, together with the map $\widehat{\pi}:= \pi\circ \kappa:J^1E\rightarrow M$, the Lagrangian density is a $\widehat{\pi}$-semibasic $d$-form on $J^1E$, expressed as
\begin{equation}\label{Eq:LagrangianDensity}
    \mathcal{L}(x^{\mu},y^a,y^a_{\mu}) = L(x^{\mu},y^a,y^a_{\mu})\,\widehat{\pi}^*\omega
\end{equation}  
in which $L:J^1E \rightarrow \mathbb{R}$ denotes the Lagrangian function. In local coordinates $(x^{\mu},y^a,y^a_{\mu})$ on $J^1E$, the volume form simply reads $\widehat{\pi}^*\omega=d^dx$. From the Lagrangian function, we define two Cartan forms $\Theta_L\in\Omega^d(J^1E)$ and $\Omega_L\in\Omega^{d+1}(J^1E)$ as
\begin{equation}\label{Eq:CartanForms}
    \Theta_L = \frac{\partial L}{\partial y^a_{\mu}}\,dy^a\wedge d^{d-1}x_{\mu} - \left(\frac{\partial L}{\partial y^a_{\mu}}\,y^a_{\mu} - L\right)\,d^dx \quad\quad\quad\quad \Omega_L := -\,d\Theta_L
\end{equation}
The quantity $d^{d-1}x_{\mu}$ is defined in the obvious manner as $d^{d-1}x_{\mu}:= \iota_{\scriptscriptstyle \partial_{\mu}} d^dx$. We refer to the pair $(J^1E,\Omega_L)$ as a Lagrangian system, further specifying this to be \textit{regular} if $\Omega_L$ is multisymplectic, and \textit{singular} if it is pre-multisymplectic \cite{de2005pre}. From the Cartan $d$-form $\Theta_L$, we define the following objects, used extensively in our analysis of scaling symmetries
\begin{equation}\label{Eq:Lagrangian1Forms}
    \theta^{\mu}_L := -\, \iota_{\partial_{d-1}}\,\cdots \;\iota_{\partial_0} (\Theta_L \wedge dx^{\mu}) = \frac{\partial L}{\partial y_{\mu}^a}\,dy^a
\end{equation}
The field equations for a Lagrangian system $(J^1E,\Omega_L)$ may be derived from a variational principle (details of which can be found in \cite{roman2009multisymplectic}); here, it suffices to note that this variational principle identifies critical sections $\phi\in\Gamma(M,E)$, whose canonical lifting $j^1\phi$ to $J^1E$ satisfy the following \textit{Euler-Lagrange field equations}
\begin{equation}\label{Eq:MultisymplecticEOM2}
    \frac{\partial}{\partial x^{\mu}}\,\left(\frac{\partial L}{\partial y^a_{\mu}}\circ j^1\phi\right) - \frac{\partial L}{\partial y^a}\circ j^1\phi = 0
\end{equation}
If $\phi$ is expressed locally as $\phi(x)=\left(x^{\mu},y^a(x)\right)$, then
\begin{equation}\label{Eq:CanonicalLifting}
    j^1\phi(x)=\left(x^{\mu},y^a(x),\frac{\partial y^a}{\partial x^{\mu}}\biggr|_x\,\right)
\end{equation}
In general, a section $\psi:M\rightarrow J^1E$ is said to be holonomic if it may be expressed as the canonical lifting $\psi=j^1\chi$ of some $\chi:M\rightarrow E$. An integrable multivector field whose integral sections all have this property is itself referred to as holonomic. The field equations may be formulated geometrically seeking locally decomposable, $\widehat{\pi}$-transverse multivector fields $\boldsymbol{X}_L$ which satisfy $\iota_{\boldsymbol{X}_L}\Omega_L = 0$. Such multivector fields are expressed locally as
\begin{equation}\label{Eq:MultivectorField}
    \boldsymbol{X}_L = f \, \bigwedge\limits_{\mu=0}^{d-1}  \left(\frac{\partial}{\partial x^{\mu}} + F_{\mu}^a\,\frac{\partial}{\partial y^a} + K_{\mu \nu}^{a}\,\frac{\partial}{\partial y^a_{\nu}}\right)
\end{equation}
for some non-vanishing $f\in C^{\infty}(J^1E)$. In practice, $\widehat{\pi}$-transversality is most easily enforced by demanding that $\iota_{\scriptscriptstyle\boldsymbol{X}}(\widehat{\pi}^*\omega)=1$, which sets $f$ to unity. When $\boldsymbol{X}_L$ is integrable, it is holonomic only when $F_{\mu}^a=y^a_{\mu}$ in the above expansion; if $F_{\mu}^a=y^a_{\mu}$, but $\boldsymbol{X}_L$ is \textit{not} integrable, then we say that it is \textit{semi-holonomic}.
\subsection{Hamiltonian Field Theory}\label{Subsec:HamiltonianFieldTheory}
When analysing systems invariant under scaling symmetries, it will be of benefit to have at our disposal both the Lagrangian and Hamiltonian descriptions. Since the contact-reduced Hamiltonian may always be obtained by performing a Legendre transform of the corresponding Lagrangian, our treatment of the Hamiltonian construction will be heavily abridged. The relevant geometrical setting is the restricted multimomentum bundle $\sigma: J^1E^*\rightarrow E$, which serves as the covariant phase space; further details on this construction may be found in \cite{echeverria2000geometry,echeverria2000multimomentum,sardanashvily1994multimomentum}. We introduce the projection $\bar{\sigma} := \pi \circ\sigma : J^1E^*\rightarrow M$, and write local coordinates on $J^1E^*$ as $(x^{\mu},y^a,p^{\mu}_a)$. The Legendre map $\mathcal{FL} : J^1E\rightarrow J^1E^*$ acts on these coordinates according to
\begin{equation}\label{Eq:LegendreMap}
    \mathcal{FL}^*\, x^{\mu} = \;x^{\mu} \quad\quad\quad\quad \mathcal{FL}^*\,y^a = y^a \quad\quad\quad\quad \mathcal{FL}^*\,p^{\mu}_a = \frac{\partial L}{\partial y^a_{\mu}}
\end{equation}
A Lagrangian system is said to be regular if $\mathcal{FL}$ is a local diffeomorphism, hyperregular if this diffeomorphism is global, and singular if $\mathcal{FL}$ is not a diffeomorphism \cite{dedecker1978problemes,dedecker2006generalization,krupka1986regular}. Throughout, we assume all systems to be (hyper)regular, and will only discuss the treatment of singular Lagrangians in our concluding remarks.\\

Application of the Legendre map to the Cartan forms (\ref{Eq:CartanForms}) gives the corresponding Hamiltonian objects
\begin{equation}\label{Eq:HamiltonCartanLocal}
    \Theta_H = p^{\mu}_a\,dy^a\wedge d^{d-1}x_{\mu} - H\,d^dx \quad\quad\quad\quad \Omega_H = -\,d\Theta_H
\end{equation}
in which $H:J^1E^*\rightarrow \mathbb{R}$ denotes the Hamiltonian function. From $\Omega_H$, we define the following 2-forms
\begin{equation}\label{Eq:Hamiltonian2-Forms}
    \omega_H^{\mu} := -\, \iota_{\partial_{d-1}}\cdots\,\iota_{\partial_0}(\Omega_H\wedge dx^{\mu}) = dy^a \wedge\,dp^{\mu}_a
\end{equation}
which will again be used extensively in section (\ref{Sec:ContactReductionHamiltonian}), and should be compared to (\ref{Eq:Lagrangian1Forms}). The equations of motion may be derived from variational arguments \cite{roman2009multisymplectic}, and if a critical section $\psi$ is expressed locally as $\psi(x) = (x^{\mu},\psi^a(x),\psi^{\mu}_a(x))$, then we find that
\begin{equation}\label{Eq:HDWEquations}
    \frac{\partial \psi^a}{\partial x^{\mu}}\biggr|_x = \frac{\partial H}{\partial p^{\mu}_a}\biggr|_{\psi(x)} \quad\quad\quad \frac{\partial \psi^{\mu}_a}{\partial x^{\mu}}\biggr|_x = -\,\frac{\partial H}{\partial y^a}\biggr|_{\psi(x)}
\end{equation}
As in the Lagrangian case, these critical sections are integral to a class of $\bar{\sigma}$-transverse, multivector fields $\{\boldsymbol{X}_H\}$, each of which satisfies $\iota_{\boldsymbol{X}_H}\Omega_H=0$.
\section{Multicontact Formulation of Classical Field Theories}\label{Sec:MulticontactFieldTheory}
As mentioned previously, the elimination of a scaling degree of freedom leaves a reduced space, which naturally inherits a multicontact structure \cite{de2023multicontact,de2025practical,rivas2025symmetries}. Physically, a theory possessing an empirically inaccessible notion of global scale is dynamically equivalent to a second, simpler theory, with no such scale, but which is frictional in nature \cite{sloan2018dynamical,sloan2021scale,bell2025dynamical,gryb2021scale}. In this section, we briefly discuss the framework required for the description of non-conservative field theories. 
\subsection{Lagrangian Formalism}\label{Subsec:MulticontactLagrangian}
Continuing to work with the fibered manifolds $\pi:E\rightarrow M$, and $\kappa:J^1E\rightarrow E$, a multicontact (or \textit{Herglotz}) Lagrangian density is a $d$-form on the space
\begin{equation}\label{Eq:MulticontactConfigurationBundle}
    \mathcal{S} := J^1E \times_M \wedge^{d-1}T^*M \cong J^1E \times \mathbb{R}^d
\end{equation}
The space $\mathcal{S}$ is of dimension $2d+n+nd$, and is a bundle over both $E$, with projection $\tau:\mathcal{S}\rightarrow E$, and $M$, with $\beta=\pi\circ\tau:\mathcal{S}\rightarrow M$. We take local coordinates on $\mathcal{S}$ to be $(x^{\mu},y^a,y^a_{\mu},s^{\mu})$, in which the $s^{\mu}$ have the field-theoretic interpretation of an action density. Pulling back the volume form $\omega\in\Omega^d(M)$ to $\mathcal{S}$, the Lagrangian density is written locally as
\begin{equation}\label{Eq:ContactLagrangian}
    \mathcal{L}(x^{\mu},y^a,y^a_{\mu},s^{\mu}) = L(x^{\mu},y^a,y^a_{\mu},s^{\mu})\,\beta^*\omega
\end{equation}
In close analogy to the multisymplectic case, we have the Lagrangian $d$-form $\Theta_L$, which, in local coordinates, is given by 
\begin{equation}\label{Eq:LagrangianForm}
    \Theta_L = \left( ds^{\mu} - \frac{\partial L}{\partial y^a_{\mu}}dy^a\right) \,\wedge \,d^{d-1}x_{\mu} + \left(\frac{\partial L}{\partial y^a_{\mu}}\,y^a_{\mu} - L\right)\,d^dx
\end{equation}
The $(d+1)$-form $\Omega_L$ is constructed from the \textit{dissipation form}, expressed in local coordinates as
\begin{equation}\label{Eq:DissipationForm}
    \sigma_{\Theta_L} = -\,\frac{\partial L}{\partial s^{\mu}}\,dx^{\mu}
\end{equation}
We then have
\begin{equation}\label{Eq:ContactOmega}
    \Omega_L = d\Theta_L +\sigma_{\Theta_L}  \wedge\,\Theta_L
\end{equation}
A holonomic section $\Psi: M\rightarrow\mathcal{S}$ may be expressed locally as 
\begin{equation*}
    \Psi(x) = \left(x^{\mu},y^a(x), \frac{\partial y^a}{\partial x^{\mu}}\biggr|_x,s^{\mu}(x)\right)
\end{equation*}
Such objects satisfy the Herglotz-Lagrange field equations
\begin{equation}\label{Eq:HerglotzLagrangeEquations}
    \begin{split}
        \frac{\partial}{\partial x^{\mu}}\left(\frac{\partial L}{\partial y^a_{\mu}}\circ \Psi\right) = \left( \frac{\partial L}{\partial y^a}+\frac{\partial L}{\partial y^a_{\mu}}\frac{\partial L}{\partial s^{\mu}}\right) \circ \Psi\hspace{2.5cm} \frac{\partial s^{\mu}}{\partial x^{\mu}} = L\circ \Psi
    \end{split}
\end{equation}
which should be considered the suitably modified version of (\ref{Eq:MultisymplecticEOM2}), adapted to accommodate the non-conservative nature of action-dependent field theories. As in the multisymplectic case, the dynamical evolution of our system may also be expressed in terms of multivector fields; in the current context, the equations of motion for locally decomposable $\beta$-transverse multivector fields $\boldsymbol{X}_L\in\mathfrak{X}^d(\mathcal{S})$ read
\begin{equation}\label{Eq:MulticontactEOM2}
    \iota_{\boldsymbol{X}_L}\Theta_L=0 \quad\quad\quad\quad\quad\iota_{\boldsymbol{X}_L}\Omega_L=0
\end{equation}
\subsection{Hamiltonian Formalism}\label{Subsec:MulticontactHamiltonian}
In parallel to the construction of $\mathcal{S}$ in (\ref{Eq:MulticontactConfigurationBundle}), we introduce
\begin{equation}\label{Eq:MultiphaseSpaceContact}
    \mathcal{S}^* := J^1E^* \times_M \wedge^{d-1}T^*M \cong J^1E^* \times \mathbb{R}^d
\end{equation}
with projection $\bar{\tau}:\mathcal{S}^*\rightarrow M$. Local coordinates on $\mathcal{S}^*$ are denoted $(x^{\mu},y^a,p^{\mu}_a,s^{\mu})$, upon which the Legendre map $\mathcal{FL}:\mathcal{S}\rightarrow \mathcal{S}^*$ acts according to
\begin{equation}\label{Eq:MulticontactLegendreMap}
    \mathcal{FL}^*\, x^{\mu} = \;x^{\mu} \quad\quad\quad\quad \mathcal{FL}^*\,y^a = y^a \quad\quad\quad\quad \mathcal{FL}^*\,p^{\mu}_a = \frac{\partial L}{\partial y_{\mu}^a} \quad\quad\quad\quad \mathcal{FL}^*\,s^{\mu}=s^{\mu}
\end{equation}
Similarly, the Hamiltonian $d$-form $\Theta_H$ on $\mathcal{S}^*$ satisfies $\mathcal{FL}^*\Theta_H=\Theta_L$, and in local coordinates, we have
\begin{equation}\label{Eq:HamiltonianForm}
    \Theta_H = \left( ds^{\mu}  - p^{\mu}_a \,dy^a\right)\,\wedge \,d^{d-1}x_{\mu} + H\,d^dx
\end{equation}
in which the Hamiltonian $H: \mathcal{S}^*\rightarrow \mathbb{R}$ is related to the Lagrangian energy function via $E_L=\mathcal{FL}^* H$. When studying the reduction of multisymplectic field theories to their frictional multicontact counterparts, it will be of benefit to work with the following 1-forms
\begin{equation}\label{Eq:Contact1Forms}
    \eta^{\mu} := -\, \iota_{\partial_m}\cdots\;\iota_{\partial_1}\,(\Theta_H\wedge dx^{\mu}) = ds^{\mu} - p^{\mu}_a\,dy^a
\end{equation}
The dissipation form for action-dependent field theories within the Hamiltonian formalism is given by
\begin{equation}\label{Eq:HamiltonianDissipationForm}
    \sigma_{\Theta_{H}} = \frac{\partial H}{\partial s^{\mu}}\,dx^{\mu}
\end{equation}
which is then used to construct the $(d+1)$-form $\Omega_{H}:=d\Theta_{H}+\sigma_{\Theta_H} \wedge\,\Theta_H$, as in (\ref{Eq:ContactOmega}). Given a holonomic section $\psi:M\rightarrow \mathcal{S}^*$, expressed as $\psi(x)=\left(x^{\mu},y^a(x),p^{\mu}_a(x),s^{\mu}(x)\right)$, the suitable modification of the Hamiltonian equations (\ref{Eq:HDWEquations}), adapted to action-dependent theories reads
\begin{equation}\label{Eq:HDWEq2}
    \frac{\partial y^a}{\partial x^{\mu}}\biggr|_x = \frac{\partial H}{\partial p^{\mu}_a}\biggr|_{\psi(x)} \quad\quad\quad \frac{\partial p^{\mu}_a}{\partial x^{\mu}}\biggr|_x = -\, \left(\frac{\partial H}{\partial y^a} + p^{\mu}_a\frac{\partial H}{\partial s^{\mu}}\right) \biggr|_{\psi(x)}\quad\quad\quad \frac{\partial s^{\mu}}{\partial x^{\mu}}\biggr|_x = \left(p^{\mu}_a\frac{ \partial H}{\partial p^{\mu}_a} - H \right)\biggr|_{\psi(x)}
\end{equation}
The final expression is simply the condition $\partial_{\mu}s^{\mu}=L^H$, expressed in canonically conjugate variables.
\section{An Introduction to Dynamical Similarity}\label{Sec:DynamicalSimilarity}
We begin by motivating the concept of scaling symmetries for simple theories of particles. In the introduction, we provided a qualitative account of how many theories are described mathematically, in a way that represents not only the evolution of observable quantities, but contains additional redundant structure, often present to reduce calculational complexity. This can be made more precise by considering the phase space of a mechanical system, where the redundancy manifests itself as a vector field, generating scale transformations that leave all dimensionless observables invariant, while acting non-trivially on the canonical coordinates. Such symmetries are \textit{not}, therefore, simple reparameterisations, which are merely alternative descriptions of the same underlying mathematical object. A scaling symmetry maps between points that are \textit{distinct} elements of the system's phase space. Such points are identified solely because all relational observables evolve in precisely the same manner. Further, unlike reparameterisation invariance in relativistic mechanics, a scaling symmetry does not have the effect of introducing constraints into the system: first-class or otherwise.\\

To continue our example from the introduction, we consider a flat, homogeneous, isotropic Friedmann model, sourced by minimally-coupled scalar fields. The phase space of this system consists of the scale factor, the Hubble rate, the scalar field values and their momenta. However, the \textit{physical observables} depend only on the Hubble rate, the scalar fields, and their velocities. The dynamical similarity within this system connects distinct points on the phase space manifold which have the same values of these observable parameters, and are thus physically indistinguishable. This is realised by a vector field, which rescales the phase space parameters in such a way so as to leave the empirically accessible ratios unchanged.\\

In order to provide a more mathematical description of these ideas, we consider a symplectic manifold $(N,\omega)$, with $\textrm{dim}\,N=2n$. Suppose that $X_H\in\mathfrak{X}^{\infty}(N)$ is the Hamiltonian vector field associated with the function $H:N\rightarrow \mathbb{R}$ via 
\begin{equation}
    \iota_{X_H}\omega = dH
\end{equation}
A vector field $D\in \mathfrak{X}^{\infty}(N)$ is said to constitute a \textit{scaling symmetry} of the Hamiltonian system $(N,\omega,H)$ if\footnote{Here, we use $\mathfrak{L}$ to denote the Lie derivative, so as to reserve $\mathcal{L}$ and $L$ for the Lagrangian density and function respectively.}
\begin{equation*}
    \mathfrak{L}_D \hspace{0.4mm}\omega = \omega \quad\quad\quad\quad\quad \mathfrak{L}_D H = \Lambda H
\end{equation*}
for some $\Lambda\in \mathbb{R}$, referred to as the \textit{degree} of the scaling symmetry. The first of these properties indicates that scaling symmetries belong to the set of non-strictly canonical transformations of $(N,\omega)$ \cite{carinena2013canonoid}. Together, these two conditions allow us to show that
\begin{equation*}
    \iota_{\scriptscriptstyle[D,X_H]}\hspace{0.3mm}\omega = [\mathfrak{L}_D,\iota_{X_H}]\hspace{0.4mm}\omega = (\Lambda-1) \,dH \quad\quad\quad\implies\quad\quad\quad [D,X_H] = (\Lambda-1) X_H
\end{equation*}
so that the vector field $D$ acts to rescale all phase space trajectories by the same non-zero factor. A non-trivial corollary of this is that there exists an algebra of invariants of the scaling symmetry, whose dynamical evolution is both autonomous and insensitive to its action \cite{sloan2018dynamical,gryb2021scale}. In this sense, a scaling symmetry maps one physical solution into a second, indistinguishable one.\\

Assuming that the flow of $D$ acts freely and properly on $N$, we have a well-defined quotient space $C:=N/\hspace{-0.2mm}\sim$, in which two points are identified if they lie on the same $D$-orbit \cite{bravetti2023scaling}. $C$ is a smooth manifold of dimension $2n-1$, and the map $\pi:N\rightarrow C$, taking points of $N$ to their equivalence class under $D$-orbits is a submersion. Further, $\xi:=\pi_*\,\textrm{ker}(\iota_D\hspace{0.2mm}\omega)$ defines a contact distribution on $C$. Locally, we identify a \textit{scaling function} $\rho:N\rightarrow \mathbb{R}$, defined to satisfy $D\rho=\rho\,$; the contact distribution $\xi$ is then represented as the kernel of the 1-form
\begin{equation}
    \pi^*\eta:=\frac{\iota_D\hspace{0.2mm}\omega}{\rho}
\end{equation}
In addition to the 1-form $\eta$, we may also introduce a contact Hamiltonian $H^c:C\rightarrow\mathbb{R}$ via
\begin{equation}
    \pi^*H^c:=\frac{H}{\rho^{\Lambda}}
\end{equation}
The system $(C,\eta,H^c)$ is a regular contact Hamiltonian system, which faithfully reproduces the dynamics of the original symplectic system, without reference to $\rho$, which, physically, we identify with the inaccessible measure of global scale, whose presence is redundant, from the perspective of the evolution of the observables.
\section{Contact Reduction of Lagrangian Systems}\label{Sec:ContactReductionLagrangian}
Despite having introduced scaling symmetries within the Hamiltonian formalism, our field-theoretic discussion commences with the (regular) Lagrangian system $(J^1E,\Theta_L)$ over the $d$-dimensional spacetime manifold $M$. In general, when a theory possesses a global measure of scale, whose presence is not required to describe the evolution of the observable degrees of freedom, there exist coordinates  $(x^{\mu},\phi,\psi^a)$ on $E$, with $a=1,\,\cdots,N$, such that under the transformation $(x^{\mu},\phi,\psi^a) \mapsto (x^{\mu}, \lambda^K \phi,\psi^a)$, the Lagrangian function is rescaled according to $L \mapsto \lambda^{\Lambda} L$ for some $\Lambda \in \mathbb{R}$. In order to preserve Poincaré invariance, we have assumed that the only dependence of $L$ on the spacetime coordinates is through that of $\phi(x)$ and $\psi^a(x)$.\\

It is relatively clear that the vector field
\begin{equation}\label{Eq:ScalingSymSigma}
    \Sigma := K\left(\phi\frac{\partial}{\partial\phi} + \phi_{\mu}\frac{\partial}{\partial\phi_{\mu}}\right) 
\end{equation}
reproduces the effect of the rescaling $(x^{\mu},\phi,\psi^a) \mapsto (x^{\mu}, \lambda^K \phi,\psi^a)$; that is $\mathfrak{L}_{\Sigma}L=\Lambda L$. It must also be verified that $\Sigma$ satisfies
\begin{equation}\label{Eq:DynamicalSimilarityCondition}
    \mathfrak{L}_{\Sigma}\,\theta_L^{\mu} = \theta^{\mu}_L \quad\quad\quad\quad\textrm{for all $\mu=0,\,\cdots\,,d-1$}
\end{equation}
in which the $\theta_L^{\mu}$ were defined in (\ref{Eq:Lagrangian1Forms}). Note that the mathematical requirement for $\Sigma$ to constitute a scaling symmetry is simply that $\mathfrak{L}_{\Sigma}\,\theta_L^{\mu} = C\theta^{\mu}_L$, for some constant $C$. Imposing that $C=1$, as above, merely fixes an overall normalisation ambiguity.\\

In order to progress with the reduction process, consider a Herglotz Lagrangian $L^H$, embedded within a multisymplectic manifold, of one dimension higher, via the expression (we omit all pullbacks via embedding maps)
\begin{equation}\label{Eq:LandLH}
    L(\rho,\rho_{\mu},\phi^a,\phi^a_{\mu}) = e^{\rho}(L^H(\phi^a,\phi^a_{\mu},s^{\mu}) + \rho_{\mu}s^{\mu})
\end{equation}
in which $\phi^a$ are a set of scalar fields, and we have made explicit all coordinate dependence. The variable $\rho$ is defined in such a way that 
\begin{equation}\label{Eq:rho}
	\rho_{\mu} := -\, \frac{\partial L^H}{\partial s^{\mu}}
\end{equation}
We shall suppose that this expression is invertible, such that the action density components $s^\mu$ may be (uniquely) expressed in terms of $\rho_\mu$, and possibly other fields. Such an assumption is not excessively restrictive, but it does allow us to express the objects on the LHS and RHS of (\ref{Eq:LandLH}) in terms of the same variables. A short calculation then shows that the equations of motion of $L$ imply the corresponding Herglotz-Lagrange field equations for $L^H$, when $L$ is restricted to the subspace upon which $L^H$ is defined. Here, the calculation is carried out expressing $s^\mu$ in terms of $\rho_\mu$ and other fields, as determined by (\ref{Eq:rho}). We thus deduce that, on this space, $L^H$ describes a dynamically equivalent theory to that of $L$, but does so without reference to $\rho$, which in this context, we interpret as the variable corresponding to an unobservable global scale.\\

In light of this, upon identifying a scaling symmetry $\Sigma$ of degree $\Lambda$, as in (\ref{Eq:ScalingSymSigma}), we define $\xi := \phi^{1/K}$, satisfying $\Sigma\,\xi=\xi$. With this, the vector field $\Sigma$ is rendered of the form
\begin{equation}\label{Eq:XiVectorField}
    \Sigma = \xi\frac{\partial}{\partial \xi} + \xi_{\mu}\frac{\partial}{\partial \xi_{\mu}}
\end{equation}
Making the identification $\xi = e^{\rho/\Lambda}$, we find that the Lagrangian adopts the following form \cite{bell2025dynamical}
\begin{equation}\label{Eq:LambdaRhoLagrangian}
    L(\rho,\rho_{\mu},\psi^a,\psi^a_{\mu}) = e^{\rho} f(\rho_{\mu},\psi^a,\psi^a_{\mu})
\end{equation}
for some function $f$ which, crucially, does \textit{not} depend upon $\rho$. In these coordinates, the scaling symmetry is simply $\Sigma=\Lambda\hspace{0.2mm}\partial_{\rho}$. The Euler-Lagrange field equation for $\rho$ implies
\begin{align*}
    0 = \frac{\partial}{\partial x^{\mu}}\left[e^{\rho} \frac{\partial f}{\partial \rho_{\mu}} \right] - e^{\rho} f = e^{\rho}\left[\rho_{\mu}\frac{\partial f}{\partial \rho_{\mu}} + \frac{\partial}{\partial x^{\mu}}\frac{\partial f}{\partial \rho_{\mu}} - f\right]\\
    \implies \quad\quad f = \rho_{\mu}\frac{\partial f}{\partial \rho_{\mu}} + \frac{\partial}{\partial x^{\mu}}\frac{\partial f}{\partial \rho_{\mu}}
\end{align*}
Comparing this expression for $f$ to the Lagrangian $L=e^{\rho}(L^H+\rho_{\mu}s^{\mu})$, and recalling that we require $\partial_{\mu}s^{\mu}=L^H$, suggests that we should identify the action density and Herglotz Lagrangian as follows
\begin{equation}\label{Eq:ActionDensityandLH}
    s^{\mu} = \frac{\partial f}{\partial \rho_{\mu}} \quad\quad\quad\quad L^H = f - \rho_{\mu}s^{\mu}
\end{equation}
Here, the first equation, defining the action density, is inverted to eliminate reference to $\rho_{\mu}$ in the function $f$ from which $L^H$ is constructed. We would now like to verify that the Herglotz-Lagrange equation
\begin{equation}\label{Eq:HerglotzEqForPsi}
        \frac{\partial}{\partial x^{\mu}}\frac{\partial L^H}{\partial \psi^a_{\mu}} - \frac{\partial L^H}{\partial \psi^a} - \frac{\partial L^H}{\partial \psi^a_{\mu}}\frac{\partial L^H}{\partial s^{\mu}} = 0
\end{equation}
faithfully reproduces the dynamics of the original system, for the unscaled fields $\psi^a$. For this, we must make careful use of the chain rule, accounting for the dependence of $s^{\mu}$ on the other coordinates. We then have
\begin{equation*}
    \begin{split}
         \frac{\partial L^H}{\partial \psi^a_{\mu}} = \frac{\partial f}{\partial \psi^a_{\mu}} + \frac{\partial f}{\partial \rho_{\nu}}\frac{\partial \rho_{\nu}}{\partial \psi^a_{\mu}} - s^{\nu}\frac{\partial \rho_{\nu}}{\partial \psi^a_{\mu}} \\
         \frac{\partial L^H}{\partial \psi^a} = \frac{\partial f}{\partial \psi^a} + \frac{\partial f}{\partial \rho_{\nu}}\frac{\partial \rho_{\nu}}{\partial \psi^a} - s^{\nu} \frac{\partial \rho_{\nu}}{\partial \psi^a}\\
         \frac{\partial L^H}{\partial s^{\mu}} = \frac{\partial f}{\partial \rho_{\nu}}\frac{\partial \rho_{\nu}}{\partial s^{\mu}} - \rho_{\mu} - s^{\nu} \frac{\partial \rho_{\nu}}{\partial s^{\mu}}
    \end{split}
\end{equation*}
Combining these results, recalling that $s^{\mu} = \partial f/\partial \rho_{\mu}$, the left-hand side of (\ref{Eq:HerglotzEqForPsi}) becomes
\begin{equation*}
    \frac{\partial}{\partial x^{\mu}}\frac{\partial f}{\partial\psi^a_{\mu}} - \frac{\partial f}{\partial \psi^a} + \rho_{\mu}\frac{\partial f}{\partial \psi^a_{\mu}}
\end{equation*}
Multiplying through by $e^{\rho}$, we recover precisely the multisymplectic field equation for $\psi^a$
\begin{equation*}
    \frac{\partial}{\partial x^{\mu}}\left(e^{\rho}\frac{\partial f}{\partial \psi^a_{\mu}}\right) - e^{\rho}\frac{\partial f}{\partial \psi^a} = 0
\end{equation*}
We have thus shown that the Herglotz Lagrangian (\ref{Eq:ActionDensityandLH}) correctly reproduces the dynamics of the original system, whilst making no reference to $\rho$, confirming that this scaling variable plays no role in the dynamical evolution of the observable degrees of freedom.
\section{Contact Reduction of Hamiltonian Systems}\label{Sec:ContactReductionHamiltonian}
Suppose that $(J^1E^*,\Omega_H)$ is a multisymplectic Hamiltonian system, with corresponding Hamiltonian function $H:J^1E^*\rightarrow \mathbb{R}$. For simplicity, we shall assume that the Legendre map is a global diffeomorphism, so that our system is hyperregular. A vector field $D\in\mathfrak{X}^{\infty}(J^1E^*)$ is said to constitute a scaling symmetry of degree $\Gamma\in\mathbb{R}$ if
\begin{equation*}
    \mathfrak{L}_D H = \Gamma H \quad\quad\quad\textrm{and}\quad\quad\quad \mathfrak{L}_D \hspace{0.3mm}\omega^{\mu}_H = \omega_H^{\mu} \quad\quad \textrm{for all}\;\mu = 0,\cdots\,,d-1
\end{equation*}
in which the 2-forms $\omega_H^{\mu}$ were defined in (\ref{Eq:Hamiltonian2-Forms}).\\

The Hamiltonian reduction may, in principle, be carried out without reference to a Lagrangian function. However, in practice, it is more convenient to work with the Hamiltonian obtained via the Legendre transform of $L=e^{\rho} f(\rho_{\mu},\psi^a,\psi^a_{\mu})$, for this implies that
\begin{equation}\label{Eq:FieldMomenta}
    \mathcal{FL}^* p^{\mu}_{\rho} = \frac{\partial L}{\partial \rho_{\mu}} = e^{\rho}\frac{\partial f}{\partial \rho_{\mu}} = e^{\rho}s^{\mu} \quad\quad\quad\quad \mathcal{FL}^*p^{\mu}_a = \frac{\partial L}{\partial \psi^a_{\mu}} = e^{\rho}\frac{\partial f}{\partial \psi^a_{\mu}}
\end{equation}
By virtue of the structure of the Lagrangian scaling symmetry $\Sigma=\partial_{\rho}$, we find that the corresponding vector field $D$ for the Hamiltonian is
\begin{equation}\label{Eq:ScalingSymmetryD}
    D = \frac{\partial}{\partial \rho} + p^{\mu}_{\rho}\frac{\partial}{\partial p^{\mu}_{\rho}}+p^{\mu}_a\frac{\partial}{\partial p^{\mu}_a}
\end{equation}
Here, the De Donder-Weyl formalism greatly facilitates our construction. The scaling symmetry \textit{must} act on both the temporal and spatial momenta; the latter do not feature amongst the phase space variables in the canonical framework. On the reduced space (the details of which will be discussed shortly) we introduce coordinates
\begin{equation}\label{Eq:ReducedCoordsHamiltonian}
    s^{\mu} = \frac{p^{\mu}_{\rho}}{e^{\rho}}\hspace{3.3cm} \Pi_a^{\mu}=\frac{p^{\mu}_{a}}{e^{\rho}}
\end{equation}
in which, for simplicity, we have omitted the pullback via the projection map. Finally, we claim that, upon defining
\begin{equation}\label{Eq:ContactHamiltonianandForms}
    H^c(\psi^a,\Pi_a^{\mu},s^{\mu}) = \frac{H}{e^{\rho}} \hspace{3.3cm} \eta^{\mu}(\psi^a,\Pi^{\nu}_a,s^{\nu}) = \frac{\iota_D\hspace{0.2mm}\omega^{\mu}_H}{e^{\rho}} = ds^{\mu} - \Pi^{\mu}_a\,d\psi^a
\end{equation}
we obtain a multicontact structure, whose corresponding Hamiltonian $d$-form $\Theta_{H^c}$ is given by
\begin{equation}\label{Eq:MulticontactReducedForm}
    \Theta_{H^c} = \eta^{\mu}\wedge\,d^{d-1}x_{\mu} + H^cd^dx
\end{equation}
Further, the contact Hamiltonian equations (\ref{Eq:HDWEq2}) for holonomic sections correctly reproduce the dynamics of the full multisymplectic system, without reference to the scaling variable $\rho$. In order to prove our claim, we shall suppose that $\gamma$ is a holonomic section, written locally as $\gamma(x) = (x^{\mu},\psi^a(x),\Pi^{\mu}_a(x),s^{\mu}(x))$; from (\ref{Eq:HDWEq2}), the equation of motion for the momenta $\Pi^{\mu}_a$ reads
\begin{align*}
    0 &= \frac{\partial\Pi^{\mu}_a}{\partial x^{\mu}} + \left(\frac{\partial H^c}{\partial \psi^a} + \Pi^{\mu}_a\frac{\partial H^c}{\partial s^{\mu}} \right)\\
    &= \frac{\partial}{\partial x^{\mu}}\left( e^{-\rho}p^{\mu}_a \right) + \left(e^{-\rho}\frac{\partial H}{\partial \psi^a} + e^{-\rho}p^{\mu}_a\frac{\partial H^c}{\partial s^{\mu}}\right)\\
    &= e^{-\rho}\left(-\rho_{\mu}p^{\mu}_a + \frac{\partial p^{\mu}_a}{\partial x^{\mu}} + \frac{\partial H}{\partial \psi^a} + p^{\mu}_a\frac{\partial H^c}{\partial s^{\mu}} \right)\\
    &= e^{-\rho}\left(\frac{\partial p^{\mu}_a}{\partial x^{\mu}} + \frac{\partial H}{\partial \psi^a}\right)
\end{align*}
In passing to the final line, we have used (\ref{Eq:rho}), noting that if $\rho_{\mu} = -\frac{\partial L^H}{\partial s^{\mu}}$ then $\rho_{\mu} = \frac{\partial H^c}{\partial s^{\mu}}$. It is thus clear that the contact Hamiltonian equation for $\Pi^{\mu}_a$ implies the corresponding multisymplectic expression for the momenta $p^{\mu}_a$. The proof for the $\psi^a$ equation of motion proceeds analogously, and is thus left as an exercise.\\

Having presented both the Lagrangian and Hamiltonian constructions of the symmetry-reduced theory, we close with a discussion of the geometry of the multicontact spaces upon which $L^H$ and $H^c$ are defined. Within the original multisymplectic Lagrangian system, we identified a symmetry of the covariant configuration space $E$, which was lifted to a vector field on the first jet bundle $J^1E$. Upon adopting coordinates along the scaling direction, we implicitly revealed that $E$ could be written as the sum of two connected pieces $E_{\pm}=(M\times\mathbb{R}_{\pm})\times_M \widetilde{E}$, in which $\widetilde{E}$ is a codimension-one subspace of $E$, identified as a configuration space comprising all unscaled field variables. $M\times\mathbb{R}_{\pm}\rightarrow M$ is a trivial bundle over $M$, and relates to a subtlety not yet discussed. Upon making a change of coordinates, the scaling vector field adopted the form (\ref{Eq:XiVectorField}); under the subsequent identification $\xi=e^{\rho/\Lambda}$, we implicitly assumed that $\xi>0$, which may not be the case. As such, in order that no dynamical information be lost, when the scaling symmetry has been rendered of the form (\ref{Eq:XiVectorField}), we should consider both $\xi=e^{\rho/\Lambda}$ \textit{and} $\xi=-\,e^{\rho/\Lambda}$. It is precisely these choices which yield the trivial bundle $M\times\mathbb{R}_{\pm}\rightarrow M$.\\

Provided both components $M\times\mathbb{R}_{\pm}$ are considered, it is clear that the quotient space under orbits of $\Sigma$ is simply $\widetilde{E}$. The reduction process eliminates only the scaling variable $\rho$, and \textit{not} its associated velocities $\rho_{\mu}$; recall that it was necessary to use the expression (\ref{Eq:ActionDensityandLH}) to eliminate reference to $\rho_{\mu}$ in favour of the action density. As a result, the reduction takes place at the level of the configuration space, and so the Herglotz Lagrangian is defined on $J^1\widetilde{E}\times \mathbb{R}^d$, in which the additional factor of $\mathbb{R}^d$ corresponds to the components of the action density, which have replaced the $\rho_{\mu}$. Within the Hamiltonian formalism, the idea is similar; however, it is now the momenta $p_{\rho}^{\mu}$ that assume the role of the action density, and so the reduced space upon which the contact Hamiltonian is defined is of the form $J^1\widetilde{E}^*\times\mathbb{R}^d$.
\section{Field Theory Classification}\label{Sec:FieldTheoryClassification}
The vast majority of physically interesting models are gauge theories \cite{taylor2001gauge,o2000gauge}; at present, our formalism is suited only to the reduction of regular theories. However, in anticipation of a need to have at our disposal a system of classification, based upon the gauge group, we shall briefly discuss the various bundle structures that arise in different kinds of theory. Our presentation follows closely that of \cite{gotay1998momentum} and \cite{gotay2004momentum}.\\

All theories we shall consider will possess a metric; however, the precise status of this metric within the theory of interest will affect the underlying multisymplectic structure. In the simplest of cases, the metric is a fixed object, so that it neither evolves dynamically, nor does it assume different values. This occurs most often when considering theories on a flat static background, so that $g_{\mu\nu}=\eta_{\mu\nu}$. Another possibility is that the metric enters the theory parametrically; in this case, we may choose to prescribe a particular background, as the context befits, but once chosen, it is fixed. For this reason, the components of the metric are included as additional variables, while the corresponding velocities are not. In practice, when a theory possesses a metric of this type, we modify the configuration space and jet bundle as follows
\begin{equation*}
    E\;\longrightarrow\; E \times_M \textrm{Sym}_2^{d-1,1}(M) \quad\quad\quad\quad\quad J^1E \;\longrightarrow\; J^1E \times_M \textrm{Sym}_2^{d-1,1}(M) 
\end{equation*}
in which $\textrm{Sym}_2^{d-1,1}(M)$ denotes the space of symmetric covariant tensors of rank two and Lorentzian signature $(d-1,1)$ on $M$; sections of $\textrm{Sym}_2^{d-1,1}(M)$ correspond to metrics on $M$. We do not consider the jet bundle over $E\times_M \textrm{Sym}_2^{d-1,1}(M)$, but append the factor of $\textrm{Sym}_2^{d-1,1}(M)$ to $J^1E$; this is precisely so that the components $g_{\mu\nu}$ of the metric are amongst the field variables, but the derivatives $g_{\mu\nu,\rho}$ are not.\\

The final possibility is that of a variational metric, which typically arises in theories coupled to gravity. As an example, consider a principal $G$-bundle $\pi: P\rightarrow M$, with gauge group $G=U(1)$; denote by $C(P)\rightarrow M$ the bundle of connections of $P$, where we may identify $C(P)\cong J^1P/G$ \cite{LopezMC2001Tgot}. Coupling this $U(1)$ gauge theory to gravity requires us to work with the following covariant configuration space 
\begin{equation*}
    E = C(P) \times_M \textrm{Sym}^{3,1}_2(M)
\end{equation*}
upon which $(x^{\mu},A_{\mu};g_{\mu\nu})$ provide local coordinates. For concreteness, we have restricted ourselves to $d=3+1$ spacetime dimensions. The gauge-invariant Lagrangian is now a function on
\begin{equation*}
    \mathcal{P} := J^1C(P)\times_M J^1(\textrm{Sym}_2^{3,1}(M))
\end{equation*}
The use of the jet bundle $J^1(\textrm{Sym}_2^{3,1}(M))$ introduces velocity variables for the components $g_{\mu\nu}$, and so local coordinates on $\mathcal{P}$ are given by $(x^{\mu},A_{\mu},g_{\mu\nu},A_{\mu,\nu},g_{\mu\nu,\rho})$; when evaluated on the jet prolongation $j^1\phi:M\rightarrow \mathcal{P}$ of some section $\phi\in \Gamma(M,E)$, the velocity coordinates become the familiar derivatives $\partial_{\nu}A_{\mu}$ and $\partial_{\rho}g_{\mu\nu}$.
\section*{Example I: N Real Scalar Fields}\label{Sec:Eg1}
As an illustration of the formalism we have developed, we consider an example of $N$ non-interacting real scalar fields $\phi^a$ (with $a=1,\,\cdots, N$) moving in a curved, $d$-dimensional spacetime $M$, equipped with a Lorentzian metric $g_{\mu\nu}$ of signature $(+ , - , -,\cdots\,)$, taken to be parametric in nature. In accordance with the discussion above, this requires us to make the replacements
\begin{equation*}
    E \;\rightarrow\; \mathcal{E} := E \times_M \textrm{Sym}^{d-1,1}_2(M)\quad\quad\quad\quad J^1E \;\rightarrow \; \mathcal{C} := J^1E \times_M \textrm{Sym}^{d-1,1}_2(M)
\end{equation*}
where $E$ is the trivial bundle $E=M\times \mathbb{R}^N \rightarrow M$, which would be an appropriate covariant configuration space for a flat background. Local coordinates on $\mathcal{C}$ are $(x^{\mu}, \phi^a, \phi^a_{\mu};g_{\mu\nu})$, where we have used a semicolon to separate the variational and parametric degrees of freedom. The Lagrangian function $L: \mathcal{C}\rightarrow\mathbb{R}$ is given by
\begin{equation}\label{Eq:NScalarLagrangian}
    L = \sqrt{-g}\sum_{a=1}^N \left[ \frac{1}{2}g^{\mu\nu} \phi_{\mu}^a\phi_{\nu}^a -\frac{1}{2}m^2(\phi^a)^2 \right]
\end{equation}
Let $X^i$ $(i=1,\,\cdots, N-1)$ denote local coordinates on $S^{N-1}$, and $n^a$ those of the embedding $n:S^{N-1}\hookrightarrow \mathbb{R}^N$. The induced metric on $S^{N-1}$, denoted $G_{ij}$, is the pullback of the flat metric $\delta_{ab}$ via the embedding $n$. Introducing a radial coordinate $R:= e^{\rho/2}\in \mathbb{R}^{\times}$, we write $\phi^a=e^{\rho/2} n^a(X)$, and the Lagrangian (\ref{Eq:NScalarLagrangian}) may be expressed as
\begin{equation}\label{Eq:NScalarRhoLagrangian}
    L = e^{\rho}\sqrt{-g}\left[g^{\mu\nu}\left(\frac{1}{8}\rho_{\mu}\rho_{\nu} + \frac{1}{2}G_{ij} X_{\mu}^i X_{\nu}^j \right) - \frac{1}{2}m^2 \right]
\end{equation}
From this Lagrangian, and making use of (\ref{Eq:Lagrangian1Forms}), we compute the 1-forms $\theta_L^{\mu}$
\begin{equation*}
    \theta_L^{\mu} = e^{\rho}\sqrt{-g}\,g^{\mu\nu}\left( \frac{1}{4}\rho_{\nu}\,d\rho + G_{ij} X^j_{\nu}\,dX^i \right)
\end{equation*}
It is then apparent that the vector field $\Sigma=\partial_{\rho}$ satisfies $\mathfrak{L}_{\Sigma}L=L$ and $\mathfrak{L}_{\Sigma}\theta_L^{\mu}=\theta_L^{\mu}$, and so is a scaling symmetry of degree one. The function $f$ used to construct the Herglotz Lagrangian in (\ref{Eq:ActionDensityandLH}) is trivially read off from (\ref{Eq:NScalarRhoLagrangian}), and we deduce the action density to be
\begin{equation*}
    s^{\mu} = \frac{\partial f}{\partial \rho_{\mu}} = \frac{1}{4}\sqrt{-g}\,g^{\mu\nu}\rho_{\nu}
\end{equation*}
Writing $L^H=f-\rho_{\mu}s^{\mu}$, and eliminating the velocity coordinate $\rho_{\mu}$, we find
\begin{equation}\label{Eq:NScalarFieldHerglotzLagrangian}
    L^H = \sqrt{-g} \left[\frac{1}{2} g^{\mu\nu}G_{ij}X^i_{\mu}X^j_{\nu} -\frac{1}{2}m^2 \right] -\frac{2}{\sqrt{-g}}g_{\mu\nu}s^{\mu}s^{\nu}
\end{equation}
Geometrically, upon introducing polar coordinates, we see that $E=(M\times \mathbb{R}^N)\backslash \{R=0\}\cong M\times (\mathbb{R}_+\times S^{N-1})$. The elimination of the scaling degree of freedom corresponds to forming the quotient space under the one-dimensional orbits of $\Sigma$. We denote this space $\widetilde{E}$, and deduce that $\widetilde{E}\cong M\times S^{N-1}$. Consequently, the Herglotz Lagrangian may be identified as a function on the space
\begin{equation}\label{Eq:NScalarReducedSpace}
    \mathcal{S} = (J^1\widetilde{E} \times_M \textrm{Sym}^{d-1,1}_2(M) )\times \mathbb{R}^d
\end{equation}
From a conservative theory of $N$ massive, non-interacting scalar fields, a change of coordinates allowed us to identify a superfluous radial scaling variable. Upon eliminating this variable, we obtain a theory of $(N-1)$ massless scalar fields subject to a constant potential, whose strength is fixed by the original mass. Further, as a consequence of the change of variables, the massless scalar fields acquire a non-trivial internal metric. Finally, the reduced theory possesses no notion of global scale, and this is compensated by an action-dependent term, which is frictional in nature. From the form of the Lagrangian, it is apparent that the massless scalar dynamics appear completely decoupled from the friction-like piece, allowing us to express (\ref{Eq:NScalarFieldHerglotzLagrangian}) as the sum $L^H=L^H_{N-1}+L^H_{\textrm{fric}}$, in which $L^H_{N-1}$ is the Lagrangian of the $(N-1)$ massless scalar fields on $J^1\widetilde{E}\times_M \textrm{Sym}^{d-1,1}_2(M)$, and $L^H_{\textrm{fric}}$ refers to the frictional component, parameterised by the action density.\\

Let us now examine the corresponding contact Hamiltonian; the most expedient way to obtain $H^c$ would be to perform a Legendre transform on (\ref{Eq:NScalarFieldHerglotzLagrangian}); however, for didactical purposes, we shall compute the full Hamiltonian from (\ref{Eq:NScalarRhoLagrangian}), and carry out the reduction using the methods of section (\ref{Sec:ContactReductionHamiltonian}). The field momenta $p^{\mu}_{\rho}$ and $p^{\mu}_i$ corresponding to $\rho$ and $X^i$ respectively are
\begin{equation}\label{Eq:NScalarFieldMomenta}
    p^{\mu}_{\rho} = \frac{\partial L}{\partial \rho_{\mu}} = \frac{1}{4}e^{\rho}\sqrt{-g}\,g^{\mu\nu}\rho_{\nu} \quad\quad\quad\quad\quad p^{\mu}_i = \frac{\partial L}{\partial X^i_{\mu}} = e^{\rho}\sqrt{-g} g^{\mu\nu}G_{ij} X_{\nu}^j
\end{equation}
Accordingly, the Hamiltonian function corresponding to (\ref{Eq:NScalarRhoLagrangian}) is given by
\begin{equation}\label{Eq:NScalarHamiltonian}
    H = \frac{1}{e^{\rho}\sqrt{-g}}g_{\mu\nu}\left( 2p^{\mu}_{\rho}p^{\nu}_{\rho} + \frac{1}{2} G^{ij} p_i^{\mu}p_j^{\nu} \right) + \frac{1}{2}e^{\rho}\sqrt{-g}\,m^2
\end{equation}
From (\ref{Eq:Hamiltonian2-Forms}), we have the 2-forms 
\begin{equation*}
    \omega_H^{\mu} = d\rho \wedge dp^{\mu}_{\rho} + dX^i \wedge dp_i^{\mu}
\end{equation*}
and, as expected, the vector field
\begin{equation}\label{NScalarVectorFieldD}
    D = \frac{\partial}{\partial \rho} + p^{\mu}_{\rho}\frac{\partial}{\partial p^{\mu}_{\rho}} +  p^{\mu}_i\frac{\partial}{\partial p^{\mu}_i}
\end{equation}
is a scaling symmetry of degree one. The contact Hamiltonian and 1-forms $\eta^{\mu}$ are easily obtained, upon introducing the following coordinates on the reduced space
\begin{equation}\label{Eq:NScalarReducedCoords}
    s^{\mu} = \frac{p^{\mu}_{\rho}}{e^{\rho}} \quad\quad\quad\quad\quad \Pi^{\mu}_i = \frac{p^{\mu}_{i}}{e^{\rho}} 
\end{equation}
We then find that
\begin{equation}\label{Eq:NScalarContactHamiltonian}
    H^c =\frac{1}{2\sqrt{-g}} g_{\mu\nu} G^{ij} \Pi^{\mu}_i\Pi^{\nu}_j + \frac{1}{2}\sqrt{-g}\,m^2 + \frac{2}{\sqrt{-g}}g_{\mu\nu}s^{\mu}s^{\nu} \quad\quad\quad\quad \eta^{\mu}  = ds^{\mu} -\Pi^{\mu}_i\,dX^i
\end{equation}
The multiphase space upon which the original Hamiltonian (\ref{Eq:NScalarHamiltonian}) was defined is $J^1E^* \times_M \textrm{Sym}_2^{d-1,1}(M)$, in which we recall that $J^1E^*$ is the multimomentum bundle over $E\cong M\times(\mathbb{R}_+\times S^{N-1})$. Having made the contact reduction, $H^c$ is now a function on
\begin{equation}\label{Eq:NScalarMultiPhaseSpace}
    \mathcal{S}^* = (J^1\widetilde{E}^* \times_M \textrm{Sym}_2^{d-1,1}(M)) \times \mathbb{R}^d
\end{equation}
where, as in (\ref{Eq:NScalarReducedSpace}), $\widetilde{E} \cong M\times S^{N-1}$. Unsurprisingly, as in the Lagrangian formalism, we observe that our contact Hamiltonian (\ref{Eq:NScalarContactHamiltonian}) consists of a piece describing a theory of $(N-1)$ massless scalar fields, with a constant shift in the energy, together with an action-dependent term, whose dissipative nature serves to compensate the inaccessibility of the global scale variable.
\section*{Example II: Interacting Scalar Fields}\label{Sec:Eg2}
Any field theory of physical interest necessarily contains interactions amongst its components; for this reason, we shall now consider a simple model of two real, interacting scalar fields $\phi$ and $\chi$, and focus our analysis more on the use of multivector fields, so as to ensure an adequate presentation of all facets of the formalism we have developed. Working over the two-dimensional spacetime manifold $M=\mathbb{R}^{1,1}$, with coordinates $(t,x)$, and fixed Minkowski metric $\eta_{\mu\nu}=\textrm{diag}(1,-1)$, we have the covariant configuration space $E=M\times \mathbb{R}_+^2$. The first jet bundle $J^1E$ admits local coordinates $(t,x,\phi,\chi,\phi_{\mu},\chi_{\mu})$, and the Lagrangian function we shall consider is 
\begin{equation}\label{Eq:SingLagrangian}
    L = \frac{1}{2}\left(\phi_t^2-\phi^2_x \right) + \frac{1}{2}\phi^2\left(\chi_t^2-\chi^2_x \right) + \phi\,(\phi_t - \phi_x)(\chi_t - \chi_x) - \chi\phi^2
\end{equation}
While this Lagrangian has mostly been chosen for its pedagogical convenience, as opposed to physical motivation, it should be noted that it is reminiscent of a non-linear sigma model, with target space metric $\textrm{diag}(1,\phi^2)$.\footnote{We are grateful to the anonymous referee for pointing this out to us.} Additionally, there are a number of dimensionful coupling constants, which, for simplicity, have been set to unity. It is relatively easy to see from (\ref{Eq:SingLagrangian}) that the vector field
\begin{equation*}
    \Sigma = \frac{1}{2}\left( \phi\frac{\partial}{\partial\phi} + \phi_t\frac{\partial}{\partial \phi_t} + \phi_x\frac{\partial}{\partial\phi_x} \right)
\end{equation*}
is a scaling symmetry of degree one. Adopting coordinates along this direction, we write $\phi \mapsto e^{\rho/2}$, with which the Lagrangian becomes
\begin{equation}\label{Eq:SingRhoLagrangian}
    L = e^{\rho}\left[\frac{1}{8}(\rho_t^2-\rho_x^2) + \frac{1}{2}(\chi_t^2-\chi_x^2) + \frac{1}{2}(\rho_t-\rho_x)(\chi_t-\chi_x)-\chi\right]
\end{equation}
A short calculation shows that the Hessian matrix for this system is given by
\begin{equation*}
    W = \frac{e^{\rho}}{4} \begin{pmatrix}
    1 & 0 & 2 & -2\\
    0 & -1 & -2 & 2\\
    2 & -2 & 4 & 0\\
    -2 & 2 & 0 & -4
    \end{pmatrix}
\end{equation*}
which is of constant rank four, confirming that $L$ is a regular Lagrangian. The corresponding multisymplectic 2-form $\Theta_L$ is given by
\begin{equation*}\label{Eq:SingTheta}
    \begin{split}
        \Theta_L = e^{\rho}\biggr[ \left(\frac{1}{4}\rho_t + \frac{1}{2} \left(\chi_t-\chi_x\right)\right) \,d\rho\wedge dx+ \left(\frac{1}{4}\rho_x +\frac{1}{2}\left(\chi_t-\chi_x\right)\right)\,d\rho\wedge dt + \left(\chi_t + \frac{1}{2}\left(\rho_t-\rho_x\right)\right)\,d\chi\wedge dx \\
        + \, \left(\chi_x + \frac{1}{2}\left(\rho_t-\rho_x\right)\right)\,d\chi\wedge dt - \left(\frac{1}{8}(\rho_t^2-\rho_x^2) + \frac{1}{2}(\chi_t^2-\chi_x^2) + \frac{1}{2}(\rho_t-\rho_x)(\chi_t-\chi_x)+\chi\right)dt\wedge dx \biggr]
    \end{split}
\end{equation*}
We shall now conduct an analysis of the dynamics described by this Lagrangian system, before carrying out the symmetry reduction. We then conclude by comparing the on-shell behaviour of each of the two theories. A straightforward computation of the Euler-Lagrange field equations for $\rho$ yields
\begin{equation}\label{Eq:SingRhoEOM}
    \frac{1}{8}\hspace{-0.5mm}\left[(\partial_t\rho)^2-(\partial_x\rho)^2 \right] - \frac{1}{2}\hspace{-0.5mm}\left[(\partial_t\chi)^2-(\partial_x\chi)^2 \right] + \frac{1}{4}(\partial_t^2-\partial_x^2)\rho + \frac{1}{2}(\partial_t-\partial_x)^2\chi +\chi =0
\end{equation}
Similarly, for $\chi$, we find that
\begin{equation}\label{Eq:SingChiEOM}
    \frac{1}{2}\left(\partial_t\rho-\partial_x\rho\right)^2 +\frac{1}{2} (\partial_t-\partial_x)^2\rho+ \partial_t\rho\,\partial_t\chi - \partial_x\rho\,\partial_x\chi + (\partial_t^2 - \partial_x^2)\chi + 1=0
\end{equation}
To illustrate how these equations of motion are deduced from a geometrical perspective, we shall assume that there exist locally decomposable $\widehat{\pi}$-transverse multivector field solutions $\boldsymbol{X}_L\in \textrm{ker}_{\widehat{\pi}}^2\,\Omega_L$, and write
\begin{equation}\label{Eq:SingMultivectorField}
    \begin{split}
        \boldsymbol{X}_L = \left(\frac{\partial}{\partial t} + F^{\rho}_t\frac{\partial}{\partial \rho} + F^{\chi}_t\frac{\partial}{\partial \chi} + G^{\rho}_{tt}\frac{\partial}{\partial \rho_t} + G^{\rho}_{tx}\frac{\partial}{\partial \rho_x} + G^{\chi}_{tt}\frac{\partial}{\partial \chi_t} + G^{\chi}_{tx}\frac{\partial}{\partial \chi_x}\right) \;\wedge \\
        \left(\frac{\partial}{\partial x} + F^{\rho}_x\frac{\partial}{\partial \rho} + F^{\chi}_x\frac{\partial}{\partial \chi} + G^{\rho}_{xt}\frac{\partial}{\partial \rho_t} + G^{\rho}_{xx}\frac{\partial}{\partial \rho_x} + G^{\chi}_{xt}\frac{\partial}{\partial \chi_t} + G^{\chi}_{xx}\frac{\partial}{\partial \chi_x}\right)
    \end{split}
\end{equation}
Here, $\widehat{\pi}$-transversality has been enforced through the condition $\iota_{\scriptscriptstyle\boldsymbol{X}_L}(dt\wedge dx)=1$, which sets the multiplicative function $f$ in (\ref{Eq:MultivectorField}) to unity. Substituting this local decomposition into the equation $\iota_{\scriptscriptstyle\boldsymbol{X}_L}\Omega_L=0$, we obtain
\begin{subequations}
    \begin{align}
        0 &= \frac{1}{8}(\rho_t^2-\rho_x^2) + \frac{1}{2}(\chi_t^2-\chi_x^2) +\frac{1}{2} (\rho_t-\rho_x)(\chi_t-\chi_x) + \chi - F^{\chi}_t \left(\chi_t + \frac{1}{2}\left(\rho_t- \rho_x\right)\right) \label{Eq:A}\\
        & +F^{\chi}_x\left(\chi_x+\frac{1}{2}\left(\rho_t-\rho_x\right)\right)+\frac{1}{4}\left( G^{\rho}_{tt} - G^{\rho}_{xx} \right) + \frac{1}{2}\left( G^{\chi}_{tt} -G^{\chi}_{tx} - G^{\chi}_{xt} +  G^{\chi}_{xx}\right)\notag\\
        \vspace{2.9mm}\notag\\
        0 &= F^{\rho}_t\left(\chi_t+\frac{1}{2}\left(\rho_t-\rho_x\right)\right) - F^{\rho}_x\left(\chi_x+ \frac{1}{2}\left(\rho_t-\rho_x\right)\right) + \frac{1}{2}\left( G^{\rho}_{tt} - G^{\rho}_{tx} - G^{\rho}_{xt} + G^{\rho}_{xx}\right)\label{Eq:B}\\
        &+G^{\chi}_{tt} -G^{\chi}_{xx}+1\notag\\
        \vspace{2.9mm}\notag\\
        0 &= \frac{1}{4}\left(\rho_t- F^{\rho}_t\right) +\frac{1}{2}\biggr[ \left(\chi_t-\chi_x\right)  - \left( F^{\chi}_t - F^{\chi}_x \right)\biggr]\label{Eq:C}\\
        \vspace{2.9mm}\notag\\
        0 &= \frac{1}{4}\left(\rho_x -F^{\rho}_x\right) + \frac{1}{2}\biggr[\left(\chi_t-\chi_x\right)  - \left(F^{\chi}_t - F^{\chi}_x\right)\biggr]\label{Eq:D}\\
        \vspace{2.9mm}\notag\\
        0 &= \left(\chi_t - F^{\chi}_t\right) + \frac{1}{2}\biggr[\left(\rho_t-\rho_x \right) - \left(F^{\rho}_t-F^{\rho}_x\right)\biggr]\label{Eq:E}\\
        \vspace{2.9mm}\notag\\
        0 &= \left(\chi_x -F^{\chi}_x\right) +\frac{1}{2} \biggr[\left(\rho_t-\rho_x\right)  - \left(F^{\rho}_t-F^{\rho}_x\right)\biggr]\label{Eq:F}
    \end{align} 
\end{subequations}
Since the Lagrangian system under consideration is regular, the self-consistency of these expressions is assured; imposing semi-holonomy of our solutions requires that we set
\begin{equation*}
    \begin{cases}
        \;\;F^{\rho}_t = \rho_t \quad\quad\quad F^{\chi}_t = \chi_t\\
        \;\;F^{\rho}_x = \rho_x \quad\quad\quad F^{\chi}_x = \chi_x
    \end{cases}
\end{equation*}
Clearly equations (\ref{Eq:C}) - (\ref{Eq:F}) are rendered trivial by this condition, whilst the first two may now be expressed as
\begin{subequations}
    \begin{align}
    0 &= \frac{1}{8}(\rho_t^2-\rho_x^2) - \frac{1}{2}(\chi_t^2-\chi_x^2) + \chi + \frac{1}{4}\left( G^{\rho}_{tt} - G^{\rho}_{xx} \right) + \frac{1}{2}\left( G^{\chi}_{tt} -G^{\chi}_{tx} - G^{\chi}_{xt} +  G^{\chi}_{xx}\right)\label{Eq:G}\\
    \vspace{3.2mm}\notag\\
    0 &= \chi_t^2-\chi_x^2 + \frac{1}{2}(\rho_t-\rho_x)(\chi_t-\chi_x) + \frac{1}{2}\left(G^{\rho}_{tt} - G^{\rho}_{tx} - G^{\rho}_{xt} + G^{\rho}_{xx}\right) +G^{\chi}_{tt} -G^{\chi}_{xx}+1\label{Eq:H}
    \end{align}
\end{subequations}
If our multivector field $\boldsymbol{X}_L=X_t\wedge X_x$ is holonomic, we require that the distribution generated by $X_t$ and $X_x$ be involutive. In such a case, we have integral sections $\psi:M\rightarrow E$, which we represent locally as $\psi(t,x) = (t,x,\rho(t,x),\chi(t,x))$; then the multivector field components $G^{\rho}_{\mu\nu}$ and $G^{\chi}_{\mu\nu}$ satisfy
\begin{equation*}
    G^{\rho}_{\mu\nu} = \frac{\partial^2 \rho}{\partial x^{\mu}\partial x^{\nu}} \quad\quad\textrm{and}\quad\quad G^{\chi}_{\mu\nu} = \frac{\partial^2 \chi}{\partial x^{\mu}\partial x^{\nu}}
\end{equation*}
Upon making these substitutions in (\ref{Eq:G}) and (\ref{Eq:H}), we recover precisely the Euler-Lagrange field equations (\ref{Eq:SingRhoEOM}) and (\ref{Eq:SingChiEOM}).\\

Having analysed the multisymplectic Lagrangian system, we may now make the symmetry reduction to obtain the Herglotz Lagrangian. From (\ref{Eq:SingRhoLagrangian}), we immediately read off the function $f(\rho_{\mu},\chi,\chi_{\mu})$, from which we obtain the following action density
\begin{equation}\label{Eq:SingActionDensity}
    s^{\mu} := \frac{\partial f}{\partial \rho_{\mu}} \quad\quad\quad\quad\quad s^t= \frac{1}{4}\rho_t + \frac{1}{2}\left(\chi_t-\chi_x \right) \quad\quad\quad s^x = -\left(\frac{1}{4}\rho_x + \frac{1}{2} \left(\chi_t - \chi_x\right)\right)
\end{equation}
Constructing the Herglotz Lagrangian $L^H=f-\rho_{\mu}s^{\mu}$, and eliminating reference to the velocity coordinates $\rho_{\mu}$, we find that
\begin{equation}\label{Eq:SingHerglotzLag}
    L^H = \frac{1}{2}(\chi_t^2-\chi_x^2) - 2\left[(s^t)^2-(s^x)^2\right] + 2(s^t+s^x)(\chi_t-\chi_x)-\chi
\end{equation}
The Herglotz-Lagrange field equations for $\chi$ then read
\begin{equation}\label{Eq:SingHerglotzEOM}
    \partial_t(\partial_t\chi + 2(s^t+s^x)) - \partial_x(\partial_x\chi + 2(s^t+s^x)) +1 = 2(\partial_t\chi-\partial_x\chi)^2 - 4(s^t\partial_t\chi+s^x\partial_x\chi) - 8(s^t+s^x)^2
\end{equation}
It is a straightforward exercise to show that, with the expressions (\ref{Eq:SingActionDensity}) for the components of the action density, this system correctly reproduces the dynamical evolution of the field $\chi$ as described by (\ref{Eq:SingRhoEOM}) and (\ref{Eq:SingChiEOM}). From this Herglotz Lagrangian, we calculate the 2-form $\Theta_{L^H}$
\begin{equation}\label{Eq:SingHerglotzTheta}
    \begin{split}
        \Theta_{L^H} = \biggr[ ds^t-\left(\chi_t + 2\left(s^t+s^x\right)\right)d\chi\biggr]\wedge \, dx \,- \,\biggr[ ds^x + \left( \chi_x +2\left(s^t+s^x\right)\right)d\chi\biggr]\wedge \,dt \\
        + \left(\frac{1}{2}(\chi_t^2-\chi_x^2) + 2\left[(s^t)^2-(s^x)^2\right]+ \chi \right) dt\wedge dx
    \end{split}
\end{equation}
together with the dissipation form
\begin{equation}\label{Eq:SingDissipationForm}
    \sigma_{\Theta} = -\,\frac{\partial L^H}{\partial s^{\mu}}\,dx^{\mu} = 2\left(2s^t-\chi_t+\chi_x\right)dt - 2\left(2s^x+\chi_t-\chi_x\right) dx
\end{equation}
The equations of motion could then be recovered geometrically; however, since this analysis is analogous to that which we have already performed, and the results of the geometric formulation are known to be equivalent to the content of the Herglotz-Lagrange field equations, we shall not pursue this further.\\

In contrast to the previous example, the presence of the interaction term $(\rho_t-\rho_x)(\chi_t-\chi_x)$ means that the reduced theory does not decouple into a piece describing the scalar dynamics, and another, independent component, containing the action density. Here, the Herglotz Lagrangian contains terms which couple $\chi$ directly to the action density, and so upon eliminating the scaling variable, the coupling between the original scalar fields manifests as an interaction between the remaining field and the frictional degrees of freedom.\\

Let us now consider the Hamiltonian formalism for this dynamical system; since our Lagrangian is regular, we know that the Legendre map is a diffeomorphism, and thus expect that the momenta be invertible, allowing us to unambiguously solve for the velocities $\rho_{\mu}$ and $\chi_{\mu}$. Indeed, we have that
\begin{equation}\label{Eq:SingRhoMomenta}
    p^{\mu}_{\rho} = \frac{\partial L}{\partial \rho_{\mu}} \quad\quad\quad\quad p^t_{\rho} = e^{\rho}\left(\frac{1}{4}\rho_t +\frac{1}{2}\left( \chi_t-\chi_x \right)\right)\quad\quad p^x_{\rho} = -\,e^{\rho}\left(\frac{1}{4}\rho_x +\frac{1}{2} \left(\chi_t-\chi_x \right)\right)
\end{equation}
and similarly for $\chi$
\begin{equation}\label{Eq:SingChiMomenta}
    p^{\mu}_{\chi} = \frac{\partial L}{\partial \chi_{\mu}} \quad\quad\quad\quad p^t_{\chi} = e^{\rho}\left(\chi_t + \frac{1}{2}\left(\rho_t 
    -\rho_x \right)\right)\quad\quad p^x_{\chi} = -\,e^{\rho}\left(\chi_x + \frac{1}{2}\left(\rho_t - \rho_x \right)\right)
\end{equation}
We then find that the Hamiltonian is given by
\begin{equation}\label{Eq:SingHamiltonian1}
    H = 2e^{-\rho} \left[ (p_{\rho}^t)^2 - (p^x_{\rho})^2 + \frac{1}{4}\left((p_{\chi}^t)^2 - (p^x_{\chi})^2 \right) -(p^t_{\rho} + p^x_{\rho})(p^t_{\chi}+p^x_{\chi}) \right] + e^{\rho}\chi
\end{equation}
As expected, the vector field
\begin{equation}\label{Eq:SingScalingSym}
    D = \frac{\partial}{\partial\rho} + p^t_{\rho}\frac{\partial}{\partial p^t_{\rho}} + p^x_{\rho}\frac{\partial}{\partial p^x_{\rho}}+ p^t_{\chi}\frac{\partial}{\partial p^t_{\chi}} + p^x_{\chi}\frac{\partial}{\partial p^x_{\chi}}
\end{equation}
constitutes a scaling symmetry of degree one; introducing the reduced space coordinates
\begin{equation*}
    s^{\mu} = \frac{p^{\mu}_{\rho}}{e^{\rho}} \quad\quad\quad\quad \Pi^{\mu}_{\chi} = \frac{p^{\mu}_{\chi}}{e^{\rho}}
\end{equation*}
we see that the contact Hamiltonian $H^c$ may be expressed as 
\begin{equation}\label{Eq:SingContactHam}
    H^c = \frac{H}{e^{\rho}} = \frac{1}{2}\left[ (\Pi^t_{\chi})^2 - (\Pi^x_{\chi})^2 \right] + \chi + 2\left[ (s^t)^2 - (s^x)^2\right] - 2(s^t+s^x)(\Pi^t_{\chi}+\Pi^x_{\chi})
\end{equation}
A straightforward calculation shows that this contact Hamiltonian is precisely the function obtained from a direct application of the Legendre map to (\ref{Eq:SingHerglotzLag}). The equations of motion derived from $H^c$ read
\begin{align*}
    \partial_t\chi &=\Pi^t_{\chi}-2(s^t+s^x) & \partial_x\chi &= -\left(\Pi_{\chi}^x + 2\left(s^t+s^x\right)\right)
\end{align*}
\vspace{-7mm}
\begin{equation}
    \partial_t\Pi^t_{\chi} + \partial_x\Pi^x_{\chi} = 2\left(\Pi^t_{\chi}+\Pi^x_{\chi}\right)^2 - 4\left(\Pi^t_{\chi}s^t - \Pi^x_{\chi}s^x \right) - 1
\end{equation}
\vspace{-7mm}
\begin{equation*}
    \partial_ts^t+\partial_xs^x = \frac{1}{2}\left[ (\Pi^t_{\chi})^2 - (\Pi^x_{\chi})^2 \right] - \chi - 2\left[ (s^t)^2 - (s^x)^2\right]
\end{equation*}
Note that, as must be the case, the final expression for $\partial_{\mu}s^{\mu}$ yields the Herglotz Lagrangian (\ref{Eq:SingHerglotzLag}), expressed in momentum variables.
\section{Singular Lagrangians and Further Applications}\label{Sec:FurtherApplications}
Before concluding, we make a number of remarks concerning singular field theories and further applicability of our formalism. Most field theories within contemporary theoretical physics possess gauge symmetries, and so are described by singular Lagrangians. The framework developed throughout is, at present, valid only for regular systems. Whilst the procedure for treating theories with gauge symmetries is, in principle, analogous to that presented here, there still remain several technical details that are unclear. For instance, given a singular multisymplectic Lagrangian $L$, one has two conceivable ways to proceed: direct contact reduction of $L$ to obtain a degenerate Herglotz Lagrangian, followed by multiphase space restriction via the geometrical algorithm \cite{de2005pre,de1996geometrical}. Alternatively, the constraint algorithm could be applied at the multisymplectic level, finding the maximal subset of $J^1E$ upon which solutions of the equations of motion exist, before seeking a scaling symmetry within the constrained theory. More work is required to ascertain the commutative relationship between these two approaches. Additionally, the geometrical interpretation of the reduced space, discussed at the end of section (\ref{Sec:ContactReductionHamiltonian}), must be suitably modified, when we are confined to work only on some subset of $J^1E$. Indeed, it must be verified that the reduced space inherits a well-defined multicontact structure; whilst this was guaranteed for regular systems, it is not, a priori, clear that the same should be true in the singular case.\\

Having ascertained the necessary  modifications of the current procedure to cases where degeneracies are present, our formalism is highly applicable to a broad array of contexts, several of which we shall now discuss. It has been demonstrated that, upon making a conformal decomposition $g_{\mu\nu}=e^{2\phi}h_{\mu\nu}$, with $\textrm{det}\, h_{\mu\nu}=-\,1$, of the spacetime metric, the Einstein-Hilbert Lagrangian admits a scaling symmetry, in which the scaling variable is precisely the conformal factor $\phi$ \cite{sloan2025dynamical}. It will be of great interest to examine the details of this dynamical similarity in the first-order vielbein formalism \cite{peldan1994actions,ashtekar1991lectures}. The use of the first-order construction is required so that we may work on the first jet bundle, which is multisymplectic; our motivation for employing the frame field and spin connection is that fermionic degrees of freedom may then be added, and the effects on the scaling symmetry examined.\\

In addition to General Relativity, many string-related low-energy effective actions are of a form in which a scaling symmetry is manifest. Consider, for example, the bosonic contribution to the NS-NS low-energy effective action of the type II theories \cite{blumenhagen2012basic} 
\begin{equation*}
    S_{\textrm{NS-NS}}=\frac{1}{2\kappa_{10}^2}\int d^{10}X\;\sqrt{-G}\,e^{-2\Phi}\left(\mathcal{R}-\frac{1}{2}|H_3|^2 + 4\,\partial_{\mu}\Phi\,\partial^{\mu}\Phi\right)
\end{equation*}
The vector field $\partial_{\Phi}$ clearly generates rescalings of the action exactly of the type discussed throughout. Of course, more work is required to ascertain whether this is a true scaling symmetry; however, there is no obstruction to our calculating the forms $\theta_L^{\mu}$, and proceeding with the formal analysis.\\

In a similar spirit, there are many low-energy descriptions of non-Abelian gauge theories coupled to dilaton-like fields. As an example, actions of the form
\begin{equation*}
    S = \int d^4x\,\left[-\frac{1}{4F(\Phi)}G_{\mu\nu}^aG^{\mu\nu}_a + \frac{1}{2}\partial_{\mu}\Phi\,\partial^{\mu}\Phi - V(\Phi) + J_a^{\mu}A^a_{\mu}\right]
\end{equation*}
have been studied extensively in \cite{chabab2007confinement} and \cite{barakat2001heavy}. For certain forms of the function $F(\Phi)$, and non-perturbative dilaton potential $V(\Phi)$, this action possesses a scaling symmetry, in which we again identify $\Phi$ as the redundant degree of freedom. Of course, the examples provided are entirely qualitative in their analysis; however, they serve to appreciate that our construction is widely applicable to many theories of current interest in high-energy particle physics and relativistic field theory.
\section{Conclusions and Outlook}\label{Sec:Conclusions}
We have advocated throughout that, while the canonical formulation has proved highly successful in the classical description, and subsequent quantisation of innumerable field theories, greater emphasis should be placed on dispensing with redundant mathematical structure. To this end, it was shown that, when described using the tools of multisymplectic geometry, many classical field theories possess degrees of freedom, corresponding to a global scale, whose presence is entirely superfluous, from the perspective of the dynamical evolution of the observables.\\

Following a brief review of the mathematical framework required to describe classical field theories in a manifestly covariant manner, we demonstrated that a multisymplectic Lagrangian (or Hamiltonian) system admitting a scaling symmetry could be reduced to a dynamically-equivalent description on a lower-dimensional space. This reduced theory was found to be action-dependent, and thus frictional in nature. Physically, we argued that this dissipative characteristic was to be interpreted as a necessary compensation in passing to a scale-insensitive description.\\ 

Through use of a pair of simple examples, we found that, in general, a theory of $N$ non-interacting massive scalar fields is dynamically-equivalent to a configuration of $(N-1)$ massless fields moving in a non-zero potential, set by the value of the original field mass, together with an independent frictional component. Further, it was observed that any coupling between the scalar fields of the original theory destroys this separation into independent pieces, and introduces terms which mix the frictional degrees of freedom with those of the remaining fields.\\

From the results presented throughout, we have discussed a number of interesting areas of future work that may be explored. For example, it is clear that the most immediate requirement is a generalisation of our construction to singular theories, where the interaction between gauge and scaling symmetries is not fully understood. Within such a framework, it will of great interest to examine the solution space of a contact-reduced description of General Relativity, particularly at those points where the conventional formalism becomes singular and thus ill-defined. It is known, for instance, that within the cosmological sector, both FLRW and class-A Bianchi models may be extended beyond the initial singularity \cite{sloan2019scalar,hoffmann2024continuation}.\\

Progress has recently been made in the construction of a multisymplectic Hamiltonian description of a simple two-dimensional supersymmetric field theory \cite{lindstrom2020covariant}. Similarly, generalised Kähler geometries have been used to construct covariant Hamiltonian formulations of non-linear sigma models \cite{lindstrom2006brief}, and so an extension of our formalism to encompass both commuting and anticommuting variables appears tractable, and will undoubtedly prove enlightening.\\

Finally, we have emphasised several times throughout the advantages presented by the covariant De Donder-Weyl formulation of classical field theories. Whilst a finite-dimensional phase space, and manifest covariance are attractive features of this approach, when contrasted with the canonical formalism, it cannot be overlooked that the latter provides a clear and rigorous path to quantisation, whereas the former does not \cite{peskin2018introduction}. In light of this, of particular interest is the development of a standardised framework, within which the restricted multimomentum bundle may be quantised, in much the same way that geometric quantisation offers a scheme for performing such calculations on the tangent bundle.
\section*{Acknowledgements}
We would like to thank the anonymous referee, whose valuable feedback contributed to the revision and improvement of this manuscript. DS also acknowledges generous support from the Foundational Questions Institute.
\bibliographystyle{unsrt}
\bibliography{Refs}
\end{document}